\documentstyle[12pt]{article}
\def\beq{\begin{equation}}
\def\eeq{\end{equation}}

\begin{document}
\begin{titlepage}
\begin{center}
{\Large \bf Theoretical Physics Institute \\ University of Minnesota \\}
\end{center}
\vspace{0.3in}
\begin{flushright}
TPI-MINN-92/57-T \\
hep-ph/9401229 \\
Revised February 1993
\end{flushright}
\vspace{0.4in}
\begin{center}
{\Large \bf Rare Decays of the Z-Boson\\}
\vspace{0.5in}
{\bf J. Jalilian-Marian\\}
School of Physics and Astronomy \\
University of Minnesota \\
Minneapolis, MN 55455 \\

\vspace{0.5in}
{\bf  Abstract  \\}
\end{center}
\vspace{.5in}

We study radiative decay modes of the Z-boson into heavy quark bound states. We
find that the widths for these decays are extremely small. We conclude that
these decays will not be detectable for the time being unless there is a
significant increase in the number of Z-bosons produced at the electron-
positron colliders.
\end{titlepage}

Rare decays of the Z-boson have recently been studied at the $e^{+}e^{-}$
colliders, CERN's LEP and the SLAC Linear Collider, where branching ratios
down to $10^{-6}$ corresponding to a few Kev can be measured $^{\cite{bearn}}$.
The partial decay widths of the
Z-boson to a photon and bound states of heavy quarks (J/$\psi$) have been
previously calculated by using Bethe-Salpeter wavefunctions for the quarkonium
state and model-dependent potentials and found to be about .2
 Kev $^{\cite{gub}}$ which might
be observebale soon. However, using a model-independent method, we
show that these decay widths are actually much smaller than previously
reported.

\vspace{.2in}\noindent
{\bf Calculation of the amplitudes A(Z$\rightarrow J/\psi\gamma$) and
A(Z$\rightarrow
\Upsilon\gamma$)\\}

We are considering the decays Z$\rightarrow J/\psi\gamma$ and Z$\rightarrow
\Upsilon\gamma$. The quark diagrams for the first decay are shown below with
similar
diagrams for the second case (replace c and $\bar{c}$ by b and $\bar{b}$).
\vspace{3in}\hfill\break
Fig.1. The quark diagrams for Z$\rightarrow J/\psi\gamma$ decay
\vspace{.2in}

Here we are assuming that the quarks inside the meson are at rest with respect
to each other. In other words, we are making the
 approximation that $P(c)=\frac{1}{2} P(J/\psi) + O(\frac{\mu}{m_{c}})$,\,where
$\mu =.5 \ Gev$ is the characteristic hadronic scale here. This is not a very
 good approximation for the c quark, but is much better for
 the b quark $^{\cite{shif}}$. Then we will have $m_{J/\psi}^2=4m^2$ ,
$(P+K)^2=M^2/2$ and
$Q=2P+K$ where $m$ and $P$ are the quark mass and momentum, $M$ and $Q$ are
 the Z-boson mass and momentum and $K$ is the photon momentum. The amplitude is
given by:

\begin{eqnarray}
A(Z\rightarrow J/\psi\gamma)&=&\epsilon^{z}_{\mu}\epsilon_{\nu}\langle J/\psi|
\bar{C}\{[(\frac{-ig}{2\cos\theta_{w}})\gamma^{\mu}(C_{v}-C_{A}\gamma^{5}) \\
\nonumber
                            & &(\frac{i}{(\not P - \not
Q)-m})(iQ_{c}e\gamma^{\nu})] +
[(iQ_{c}e\gamma^{\nu})(\frac{i}{(\not Q - \not P)-m}) \\  \nonumber
                            &
&(\frac{-ig}{2\cos\theta_{w}})\gamma^{\mu}(C_{v}-C_{A}\gamma^{5})]\}
 C|0\rangle
\end{eqnarray}

\vspace{.1in}
Here, $\epsilon^{z}_{\mu}$ and $\epsilon_{\nu}$ are polarization vectors of the
$Z$-boson and photon
respectively and $e^{2}=4\pi\alpha$ where $\alpha = 1/137$ is the
electromagnetic coupling constant
and $g=\frac{e}{\sin\theta_{w}}$,\, $Q_{c}=2/3$ is the charmed quark electric
charge,\,
 $C_{v}=T^{3} - 2\sin^{2}\theta_{w} Q_{c}$ and $C_{A}=T^{3}$ are respectively
the
 vector and axial vector coupling in the Weinberg-Salam theory,\,$T^{3}$ \ is
 equal to $\frac{1}{2}$
 for
 a c quark and $\frac{-1}{2}$ for a b quark, \ $\sin^{2}\theta_{w}\simeq .23$
and
 $\sigma^{\mu\nu}=i/2 (\gamma^{\mu}\gamma^{\nu} -
 \gamma^{\nu}\gamma^{\mu})$.

Eq.(1) can be simplified, after some $\gamma$ matrix algebra and using the
 identity $\gamma^{\mu}\gamma^{\lambda}\gamma^{\nu}=g^{\mu\lambda}\gamma^{\nu}
+
 g^{\lambda\nu}\gamma^{\mu} - g^{\mu\nu}\gamma^{\lambda} + i\epsilon^{\sigma
\mu\lambda\nu}\gamma_{\sigma}\gamma^{5}$ and the Dirac equation $\bar{U}(P)
(\not P -m)=0$ and $(\not P +m)V(P)=0$ to,

\begin{eqnarray}
A(Z\rightarrow J/\psi\gamma)& \simeq & \frac{2iQ_{c}eg}{3M^{2}\cos\theta}
 \epsilon^{z}_{\mu}\epsilon_{\nu}\langle J/\psi|\bar{C}\{[-2iC_{v}
                                        (Q_{\lambda}-P_{\lambda})
\epsilon^{\sigma\mu\lambda\nu}\gamma_{\sigma}\gamma^{5}]- \\  \nonumber
                            &        &[C_{A}(4mi\sigma^{\nu\mu}\gamma^{5}-
                                         2P^{\mu}\gamma^{\nu}\gamma^{5} +
                                       2P^{\nu}\gamma^{\mu}\gamma^{5} + \\
\nonumber
                            &        &
2i\epsilon^{\sigma\mu\lambda\nu}Q_{\lambda}
                                        \gamma_{\sigma})]\}C|0\rangle
\end{eqnarray}

\vspace{.1in}
The term involving $\sigma^{\nu\mu}\gamma^{5}$ is proportional to
$\epsilon^{\nu\mu\lambda\sigma}\sigma_{\lambda\sigma}$ and in general, has
non-zero matrix element in the above expression, but its' contribution
is of the order of  $\frac{m^{2}}{M^{2}}$ and here, we have neglected
$\frac{m^{2}}{M^{2}}$ terms.
Now, using the fact that the matrix element of a  pseudo-vector
taken between a vector state and vacuum state is zero, we get,

\beq
\ A(Z\rightarrow J/\psi\gamma)\simeq \frac{-4Q_{c}egC_{A}}{3M^{2}\cos\theta}
\epsilon^{z}_{\mu}\epsilon_{\nu}\epsilon^{\sigma\mu\lambda\nu}
Q_{\lambda}
\langle J/\psi|\bar{C}\gamma_{\sigma}C|0\rangle
\eeq

The matrix element of the vector current in the above equation can be written
 as,

\beq
\langle J/\psi|\bar{C}\gamma_{\sigma}C|0\rangle=4g_{v}m^{2}\phi_{\sigma}
\eeq
where $\phi_{\sigma}$
 is the polarization vector of J/$\psi$ and $g_{v}$ is experimentally found
 to be about $.13$. Now, squaring the amplitude will give us,

\[ | \bar{A} (Z \rightarrow J/\psi\gamma)|^{2}\simeq \frac{224\pi^{2}\alpha^{2}
Q^{2}_{c}C^{2}_{v}g^{2}_{v}m^{4}}{\cos^{2}\theta_{w}\sin^{2}\theta_{w} M^{2}}
\]

For the (Z$\rightarrow \Upsilon\gamma)$ decay,
 we follow the same procedure with $\langle
\Upsilon|\bar{b}\gamma^{\mu}b|0\rangle =
 4g\prime_{v}m^{2}_{b}\phi^{\mu}$
and $g\prime_{v}$ is found from $\Upsilon\rightarrow e^{+}e^{-}$
 decay rates to be about $.04$ \ . Then,

\[ | \bar{A} (Z \rightarrow \Upsilon\gamma)|^{2} \simeq
 \frac{224\pi^{2}\alpha^{2}Q^{2}_{b}C^{2}_{v}g\prime^{2}_{v}m^{4}_{b}}
{\cos^{2}\theta_{w}\sin^{2}\theta_{w} M^{2}} \]
and for the decay rates, using $M=91 Gev,m_{c}\simeq 1.5 Gev, m_{b}\simeq 4.5
Gev$ and massless two-body
decay phase space, we get $\Gamma (Z \rightarrow J/\psi\gamma)\sim  10^{-10}
Gev$
and
$\Gamma (Z \rightarrow \Upsilon\gamma)\sim  10^{-9} Gev$

To summarize, we have shown that decay widths for Z$\rightarrow J/\psi\gamma$
  and Z$\rightarrow\Upsilon\gamma$ are much smaller
 than previously reported and unfortunately, will not be
experimentally observable for the time being.

I would like to thank Professor
M.A. Shifman for bringing this problem to my attention and many helpful
discussions. I would also like to thank the referee of Zeitschrift f\"{u}r
Physik C for pointing out a most crucial error in the earlier version of this
paper.


\begin{thebibliography} {99}
\bibitem{bearn}
 W. Bearnreuther, M.J. Duncan, E.W.N. Glover, R. Kleiss, J.J. Van der Bij,
J.J. Gomez-Cadenas and C.A. Heusch, CERN Report No. TH.5484/89 (unpublished);
T.J. Weiler, in Proceedings of the Second International Symposium on the
 Fourth
Family of Quarks and Leptons, Santa Monica, California, 1989, edited by D.B.
Cline and A. Soni (New York Academy of Sciences, New York).

\bibitem{gub}
G. Guberina, J.H. K\"{u}hn, R.D. Peccei, and R. R\"{u}ckl, Nucl. Phys. B174,
317(1980); J.H. K\"{u}hn, Acta. Phys. Pol. B12, 347(1981).

\bibitem{shif}
M.A. Shifman and M.I. Vysotsky, Nucl. Phys. B186, 475(1981).
\end{thebibliography}
\end{document}